\documentclass[twocolumn, aps, showpacs]{revtex4}
\usepackage{graphicx}
\begin{document}
\title{Electron-acoustic plasma waves: \\ oblique modulation and
envelope solitons \footnote{Preprint; submitted to
\textit{Physical Review E}.}}
\author{I. Kourakis}
\altaffiliation[On leave from: ]{U.L.B. - Universit\'e Libre de
Bruxelles, Physique Statistique et Plasmas C. P. 231, Boulevard du
Triomphe, B-1050 Brussels, Belgium; also: U. L. B., Facult\'e des
Sciences Apliqu\'ees - C.P. 165/81 Physique G\'en\'erale, Avenue
F. D. Roosevelt 49, B-1050 Brussels, Belgium}
\email{ioannis@tp4.rub.de}
\author{P. K. Shukla}
\email{ps@tp4.rub.de}
\affiliation{Institut f\"ur Theoretische Physik IV, Fakult\"at
f\"ur Physik und Astronomie, Ruhr--Universit\"at Bochum, D--44780
Bochum, Germany}

\date{\today}

\begin{abstract}
Theoretical and numerical studies are presented of the amplitude
modulation of electron-acoustic waves (EAWs) propagating in
space plasmas whose constituents are inertial cold electrons,
Boltzmann distributed hot electrons and stationary ions.
Perturbations oblique to the carrier EAW propagation direction
have been considered. The stability analysis, based on a nonlinear
Schr\"odinger equation (NLSE), reveals that the EAW may become
unstable; the stability criteria depend on the angle $\theta$ between
the modulation and propagation directions. Different types of
localized EA excitations are shown to exist.
\end{abstract}



\maketitle


\section{Introduction}

Electron acoustic waves (EAWs) are high-frequency (in comparison with
the ion plasma frequency) electrostatic modes \cite{STB} in plasmas
where a `minority' of inertial cold electrons oscillate against a
dominant thermalized background of inertialess hot electrons providing the
necessary restoring force. The phase speed $v_{ph}$ of the EAW
is much larger than the thermal speeds of cold electrons ($v_{th, c}$) and ions
($v_{th, i}$), but is much smaller than the thermal speed of the hot electron
component ($v_{th, h}$); $v_{th, \alpha} = (T_\alpha/m_\alpha)^{1/2}$, where
$\alpha = c, h, i$ ($m_\alpha$ denotes the mass of the component $\alpha$;
the Boltzmann constant $k_B$ is understood to precede the temperature $T_\alpha$
everywhere).  Thus, the ions may be regarded as a constant positive charge density
background, providing charge neutrality. The EAW frequency is typically well
below the cold electron plasma frequency, since the wavelength is larger than the
Debye length $\lambda_h =(T_h/4\pi n_{h0}e^2)^{1/2}$ involving hot electrons
($n_\alpha$ denotes the particle density of the component $\alpha$ everywhere).

The linear properties of the EA waves are well understood [2--5].
Of particular importance is the fact that the EAW propagation
is only possible within a restricted range of the parameter values, since both long
and short wavelength EAWs are subject to strong Landau damping due to
resonance with either the hot or the cold (respectively) electron component.
In general, the EAW group velocity scales as $v_{ph} = v_{th, h} \,
\sqrt{n_c/n_h}$; therefore, the condition $v_{th, c} \ll v_{ph} \ll v_{th, h}$
immediately leads to a stability criterion in the form: $T_c/T_h \ll \sqrt{n_c/n_h} \ll 1$.
A more rigorous investigation [4] reveals that EAWs will be heavily damped,
unless the following (approximate) conditions are satisfied: $T_h/T_c \gtrsim 10$ and
$0.2 \lesssim n_c/n_e \lesssim 0.8$ (where $n_e = n_c + n_h$).  Even then, however, only
wavenumbers $k$ between, roughly, $0.2 \, k_{D, c}$ and $0.6 \, k_{D, c}$ (for $T_h/T_c = 100$;
see in Ref. 4b), will remain weakly damped [$k_{D, c} = (4\pi n_{c, 0}e^2/T_c)^{1/2}
\equiv \lambda_{D, c}^{-1}$ obviously denotes the cold electron Debye wavenumber].
The stable wavenumber value range is in principle somewhat extended with
growing temperature ratio $T_h/T_c$; see the exhaustive discussion in Refs. 4 and 5.

As far as the {\em{nonlinear}} aspects of EAW are concerned,
the formation of coherent EA structures has been considered in a
one-dimensional model involving cold \cite{Dubouloz, singh} or
finite temperature \cite{Mace2} ions. Furthermore, such
non-envelope solitary structures, associated with a localized
compression of the cold electron density, have been shown to exist
in a magnetized plasma [9 -- 11]. It is worth noting that such
studies are recently encouraged by the observation of moving
EAW-related structures, reported by spacecraft missions e.g. the
FAST at the auroral region [12 -- 14], as well as the GEOTAIL and
POLAR earlier missions in the magnetosphere [15 -- 17].  However,
although most solitary wave structures observed are positive
potential waves (consistent with an electron hole image), there
have also been some negative potential and low velocity structure
observations, suggesting that some other type of solitary waves
may be present in the magnetosphere [17, 18].  These structures
are now believed to be related to EA envelope solitary waves, for
instance due to trapping and modulation by ion acoustic density
perturbations \cite{Dubouloz, shukla}.

Amplitude modulation is a long-known generic feature of nonlinear
wave propagation, resulting in higher harmonic generation due to
nonlinear self-interactions of the carrier wave in the background
medium. The standard method for studying this mechanism is a
multiple space and time scale technique \cite{redpert, redpert2},
which leads to a nonlinear Schr\"odinger equation (NLSE)
describing the evolution of the wave envelope. Under certain
conditions, it has been shown that waves may develop a
Benjamin-Feir-type (modulational) instability (MI), i.e. their
modulated envelope may be unstable to external perturbations.
Furthermore, the NLSE-based analysis, encountered in a variety of
physical systems [21 -- 23], reveals the possibility of the
existence of localized structures (envelope solitary waves) due to
the balance between the wave dispersion and nonlinearities. This
approach has long been considered with respect to electrostatic
plasma waves [20, 24 -- 30].

In this paper, we study the occurrence of modulational instability
as well as the existence of envelope solitary structures involving
EAWs that propagate in an unmagnetized plasma composed of three
distinct particle populations: a population of `cold' inertial
electrons (mass $m_e$, charge $- e$), surrounded by an environment
of `hot' (thermalized Boltzmann) electrons, moving against a fixed
background of ions (mass $m_i$, charge $q_i = + Z_i e$), which
provide charge neutrality. These three plasma species will
henceforth be denoted by $c,\, h$ and $i$, respectively. By
employing the reductive perturbation method and accounting for
harmonic generation nonlinearities, we derive a cubic
Schr\"odinger equation for the modulated EA wave packet. It is
found that the EAWs are unstable against oblique modulations.
Conditions under which the modulational instability occurs are
given. Possible stationary solutions of the nonlinear
Schr\"odinger equation are also presented.

\section{The model equations}

Let us consider the hydrodynamic--Poisson system of equations for the
EAWs in an unmagnetized plasma. The number density $n_c$ of cold
electrons is governed by the continuity equation
\begin{equation}
\frac{\partial n_c}{dt} + \nabla \cdot (n_c \,\mathbf{u}_c)= 0 \, ,
\label{densityequation}
\end{equation}
where the mean velocity $\mathbf{u}_c$ obeys
\begin{equation}
\frac{\partial \mathbf{u}_c}{dt} + \mathbf{u}_c \cdot \nabla
\mathbf{u}_c \, = \, \frac{e}{m_e}\,\nabla \,\Phi  \, .
\end{equation}
Here, the wave potential $\Phi$ is obtained from Poisson's equation
\begin{equation}
\nabla^2 \Phi \, =\, - 4 \pi \, \sum q_s\, n_s \, = \,4 \pi \,e
\,(n_c  \,+ n_h  \, - Z_i \,n_i)  \, .
\label{Poisson}
\end{equation}
We assume immobile ions ($n_i =n_{i,0}$ = constant) and a Boltzmann distribution
for the hot electrons, i.e. \(n_h \approx n_{h,0}\, \exp(e \Phi/k_B T_h) \, \)
($T_h$ is electron temperature, $k_B$ is the Boltzmann constant),
since the EAW frequency is much higher than the ion plasma frequency,
and the EA wave phase velocity is much lower than the electron thermal speed
$(T_h/m_e)^{1/2}$.  The overall quasi-neutrality condition reads
\begin{equation}
n_{c,0} \, + n_{h, 0} \, - Z_i \, n_{i, 0} \, =\, 0 \, .
\label{neutrality}
\end{equation}


Re-scaling all variables and developing around $\Phi = 0$, Eqs.
(\ref{densityequation}) - (\ref{Poisson}) can be cast in the
reduced form
\begin{eqnarray}
\frac{\partial n}{dt} + \nabla \cdot (n \,\mathbf{u})&= & 0\, ,
\nonumber \\
\frac{\partial \mathbf{u}}{dt} + \mathbf{u}  \cdot \nabla
\mathbf{u} \, &=& \,\nabla \phi
\, ,\nonumber
\end{eqnarray}
and
\begin{equation}
\nabla^2 \phi \, =\, \phi + \frac{1}{2} \,\phi^2 + \frac{1}{6}
\,\phi^3 + \,\beta\,(n - 1)\, , \label{reducedeqs}
\end{equation}
where all quantities are non-dimensional: $n = n_c/n_{c, 0}$,
$\mathbf{u} = \mathbf{u}_c/v_{0}$ and $\phi = \Phi/\Phi_0$; the
scaling quantities are, respectively: the equilibrium density
$n_{c, 0}$, the `electron acoustic speed' $v_{0} = c_{s, h} = (k_B
T_h/m_e)^{1/2}$ and $\Phi_{0} = (k_B T_h/e)$. Space and time are
scaled over the Debye length $\lambda_{D, h} = (k_B T_h/4 \pi
n_{h, 0} e^2)^{1/2}$ and the inverse plasma frequency $\omega_{p,
h}^{-1} = \lambda_{D, h}/c_s = (4 \pi n_{h, 0} e^2/m_e)^{- 1/2}$,
respectively. The dimensionless parameter $\beta$ denotes the
ratio of the cold to the hot electron component i.e. \( \beta =
n_c/n_h \). Recall that Landau damping in principle prevails on
both high and low values of $beta$ (cf. the discussion in the
introduction). According to the results of Ref. [4b], for undamped
EA wave propagation one should consider: $0.25 \lesssim \beta
\lesssim 4$.

\section{Perturbative analysis}

Let $S$ be the state (column) vector $(n, \, \mathbf{u} , \,
\phi)^T$, describing the system's state at a given position
$\mathbf{r}$ and instant $t$. Small deviations will be considered
from the equilibrium state $S^{(0)} = (1, \, \mathbf{0} , \,0)^T$
by taking \( S = S^{(0)} \, + \, \epsilon \, S^{(1)} + \,
\epsilon^2 \, S^{(2)} + \, ... = S^{(0)} \, + \, \,
\sum_{n=1}^\infty \epsilon^n \, S^{(n)}\),  where $\epsilon \ll 1$
is a smallness parameter. Following the standard multiple scale
(reductive perturbation) technique \cite{redpert}, we shall
consider a set of stretched (slow) space and time variables \(
\zeta \,= \, \epsilon (x - \lambda \,t) \) and \( \tau \,= \,
\epsilon^2 \, t \, , \label{slowvar} \), where $\lambda$ is to be
later determined by compatibility requirements. All perturbed states
depend on the fast scales via the phase
$\theta_1 = \mathbf{k \cdot r} - \omega t$ only, while the slow
scales only enter the $l-$th harmonic
amplitude $S_l^{(n)}$, viz. \( S^{(n)} \,= \,
\sum_{l=-\infty}^\infty \,S_l^{(n)}(\zeta, \, \tau)
 \, e^{i l (\mathbf{k \cdot r} - \omega t)}
\); the reality condition $S_{-l}^{(n)} = {S_l^{(n)}}^*$ is met by
all state variables. Two directions are therefore of importance in
this (three-dimensional) problem: the (arbitrary) propagation
direction and the oblique modulation direction, defining, say, the
$x-$axis, characterized by a pitch angle $\theta$. The wave
vector $\mathbf{k}$ is thus taken to be \( \mathbf{k} = (k_x, \,
k_y) = (k\, \cos\theta, \, k\, \sin\theta) \).

Substituting the above expressions into the system of equations
(\ref{reducedeqs}) and isolating distinct orders in $\epsilon$, we
obtain the $n$th-order reduced equations
\begin{eqnarray}
- i l \omega n_l^{(n)} \,+\, i l \mathbf{k \cdot u}_l^{(n)} 
\,-\,
\lambda \, \frac{\partial n_l^{(n-1)}}{\partial \zeta}
\,+\,
\frac{\partial n_l^{(n-2)}}{\partial \tau}
\nonumber \\
\,+\,
\frac{\partial u_{l, x}^{(n-1)}}{\partial \zeta} 
\nonumber \\
\,+\, \sum_{n' = 1}^{\infty} \, \sum_{l' = -\infty}^{\infty}
\biggl[  i l \mathbf{k \cdot u}_{l-l'}^{(n-n')} \, n_{l'}^{(n')} 
\qquad \qquad \qquad 
\nonumber \\ 
+\,
\frac{\partial}{\partial \zeta}
\biggl( n_{l'}^{(n')} u_{(l-l'), x}^{(n-n'-1)}\biggr)
\biggr]  \,
= \, 0 \, , \quad
\label{geneqn1} \\
\nonumber \\
- i l \omega \mathbf{u}_l^{(n)} \,- \, i l \mathbf{k} \phi_l^{(n)}
\,-\, \lambda \, \frac{\partial \mathbf{u}_l^{(n-1)}}{\partial
\zeta} \,+\, \frac{\partial \mathbf{u}_l^{(n-2)}}{\partial \tau}
\nonumber \\
\,-\, \frac{\partial \phi_{l}^{(n-1)}}{\partial \zeta} \,\hat x
\nonumber \\
\,+\, \sum_{n' = 1}^{\infty} \, \sum_{l' = -\infty}^{\infty}
\biggl[  i l' \mathbf{k \cdot u}_{l-l'}^{(n-n')} \, \mathbf{u}_{l'}^{(n')} 
\qquad \qquad \qquad \nonumber \\ +\,
 u_{(l-l'), x}^{(n-n'-1)}\,
\frac{\partial \mathbf{u}_{l'}^{(n')}}{\partial \zeta} \biggr] \,
= \, 0 \, ,\label{geneqn2}
\end{eqnarray}
and
\begin{eqnarray}
- (l^2 k^2 + 1)\, \phi_l^{(n)} \,-\,\beta \, n_l^{(n)} 
\qquad \qquad \qquad \qquad \qquad \qquad 
\nonumber  \\
\,  + 2 i l
k_x \, \frac{\partial \phi_l^{(n-1)}}{\partial \zeta} \,+\,
\frac{\partial^2 \phi_l^{(n-2)}}{\partial \zeta^2} \qquad \qquad 
\qquad \qquad 
\nonumber \\
-\, \frac{1}{2} \,\sum_{n' = 1}^{\infty} \, \sum_{l' =
-\infty}^{\infty} \, \phi_{l-l'}^{(n-n')} \,\phi_{l'}^{(n')} \,
\qquad \qquad \qquad \qquad 
\nonumber \\
-\, \frac{1}{6}\,\sum_{n', n'' = 1}^{\infty} \, \sum_{l', l'' =
-\infty}^{\infty} \, \phi_{l-l'-l''}^{(n-n'-n'')}
\,\phi_{l'}^{(n')}\,\phi_{l''}^{(n'')}
 \, = \, 0 \, . \quad
\label{geneqn4}
\end{eqnarray}
For convenience, one may consider instead of the vectorial
relation (\ref{geneqn2}) the scalar one obtained by taking its scalar
product with the wavenumber $\mathbf{k}$.

The standard perturbation procedure now consists in solving in
successive orders $\sim \epsilon^n$ and substituting in subsequent
orders. For instance, the equations for $n = 2$, $l = 1$
\begin{eqnarray}
- i l \omega n_l^{(1)} \,+\, i l \mathbf{k \cdot u}_l^{(1)} \, = \, 0 \, ,
\label{1eqn1}
\\
- i l \omega \mathbf{u}_l^{(1)} \,- \, i l \mathbf{k} \phi_l^{(1)}
\, = \, 0 \label{1eqn2}
\end{eqnarray}
and
\begin{equation}
- (l^2 k^2 + 1)\, \phi_l^{(1)} \,-\,\beta \, n_l^{(1)} \, = \, 0
\label{1eqn4}
\end{equation}
provide the familiar EAW dispersion relation
\begin{equation}
\omega^2\,  = \frac{\beta \, k^2}{k^2 + 1} \, ,
\label{dispersion}
\end{equation}
i.e. restoring dimensions
\begin{eqnarray}
\omega^2\,  & = & \omega_{p, c}^2\, \frac{k^2}{k^2 + k_{D}^2 } \,
\equiv \,\frac{c_{s, c}^2\, k^2}{1 + k^2 \, {\lambda_{D}}_{h}^2} \, ,
\label{dispersion-dim}
\end{eqnarray}
where $\omega_{p, c} = c_{s, c}/\lambda_{D, c} =
(4 \pi n_{c, 0} e^2/m_e)^{1/2}$ (associated with the cold component),
and determine the first harmonics of the perturbation viz.
\begin{eqnarray}
n_1^{(1)} \,  & = & -\, \frac{1 + k^2}{\beta} \phi_1^{(1)}
\, , \qquad \mathbf{k\cdot u}_1^{(1)}\, =\, \omega \, n_1^{(1)} \,
\nonumber \\
\qquad u_{1, x}^{(1)} \, & = & \frac{\omega}{k} \cos\theta \,
n_1^{(1)} \,
\, , \qquad  u_{1, y}^{(1)}  \, = \frac{\omega}{k} \sin\theta \,
n_1^{(1)}
\, \, . \qquad \label{coeffs11}
\end{eqnarray}

Proceeding in the same manner, we obtain the second order quantities,
namely the amplitudes of the second harmonics $S_2^{(2)}$
and constant (`direct current') terms $S_0^{(2)}$,
as well as a non-vanishing contribution $S_1^{(2)}$ to the first harmonics.
The lengthy expressions for these
quantities, omitted here for brevity, are conveniently expressed in terms
of the first-order potential correction $\phi_1^{(1)}$.
The equations for $n = 2$, $l=1$
then provide the compatibility condition: \( \lambda \,  = v_g(k) \,  =
\frac{\partial \omega}{\partial k_x} = \omega'(k) \cos\theta =$ $
\frac{\omega^3}{\beta k^3} \,\cos\theta\); $\lambda$ is, therefore,
the group velocity in the $x$ direction.

\section{Derivation of the Nonlinear Schr\"odinger Equation}

Proceeding to the third order in $\epsilon$ ($n=3$), the equations for
$l = 1$ yield an explicit compatibility condition in the form of
the Nonlinear Schr\"odinger Equation
\begin{equation}
i\, \frac{\partial \psi}{\partial \tau} + P\, \frac{\partial^2 \psi}{\partial\zeta^2}
+ Q \, |\psi|^2\,\psi = 0
\, .
\label{NLSE}
\end{equation}
where $\psi$ denotes the electric potential correction $\phi_1^{(1)}$.
The `slow' variables $\{ \zeta, \tau \}$ were defined above.

The {\em group dispersion coefficient} $P$ is related to the curvature
of the dispersion curve as \( P \,  = \, \frac{1}{2} \,
\frac{\partial^2 \omega}{\partial k_x^2} \,= \, \frac{1}{2}\,
\biggl[ \omega''(k) \, \cos^2\theta \, + \omega'(k) \,
\frac{\sin^2\theta}{k} \biggr] \); the exact form of P reads
\begin{equation}
P(k) \,  =\, \frac{1}{\beta}
\frac{1}{2\,\omega} \, \biggl( \frac{\omega}{k}\biggr)^4\,
\biggl[ 1 - \biggl(1 + 3\, \frac{1}{\beta} \,\omega^2 \biggr)\, \cos^2\theta
\biggr] \, .
\label{Pcoeff}
\end{equation}

The {\em nonlinearity coefficient} $Q$ is due to the carrier wave
self-interaction in the background plasma. Distinguishing different
contributions, $Q$ can be split into three distinct parts, viz.
\( Q = \, Q_0 \, +\, Q_1 \, +\, Q_2 \), where
\begin{eqnarray}
Q_0 &=& \, \frac{\omega^3}{2 \,\beta^3 \,k^2}\, \frac{1}{(1+
\,k^2)^3 - \cos^2\theta}\, 
\nonumber
\\ & &
\biggl\{ (1+ \,k^2)^4 (1+ 2 \beta +
\,k^2) \nonumber
\\ & &
+ \biggl[ \beta^2 + 4 \beta (1+ \,k^2)^3 + 4 (1+ \,k^2)^4 (2 + \,2
k^2+ \, k^4) \biggr] \, \nonumber
\\ & & \qquad \times \cos^2\theta \biggr\}
 \, , \label{Q0coeff}
\\ Q_1 &=&
\, \frac{\omega^3}{4 \,\beta \,k^2} \, ,\label{Q1coeff}
\\
Q_2 &= & \, - \frac{k^2}{12\, \omega^3} \, \biggl[
\frac{\omega^6}{\beta \,k^6} \,+ \frac{\omega^2}{k^2} \,+ 3 \, (3
+ 8 \, k^2 ) \biggr] \, .
 \label{Q2coeff}
\end{eqnarray}
We observe that only the first contribution $Q_0$, related to
self-interaction due to the zeroth harmonics, is angle-dependent,
while the latter two - respectively due to the cubic and quadratic
terms in (\ref{reducedeqs}c) - are \emph{isotropic}. Also, $Q_2$
is negative, while $Q_0$, $Q_1$ are positive for all values of $k$
and $\beta$. For parallel modulation, i.e. $\theta = 0$, the
simplified expressions for $\left.P\right|_{\theta=0}$ and
$\left.Q\right|_{\theta=0}$ are readily obtained from the above
formulae; note that $\left.P\right|_{\theta=0} < 0$, while
$\left.Q\right|_{\theta=0}$, even though positive for $k
\rightarrow 0$ (see below), changes sign at some critical value of
$k$.


A preliminary result regarding the behaviour of the
coefficients $P$ and $Q$ for long wavelengths may
be obtained by considering the limit of small $k \ll 1$ in the
above formulae.  The parallel ($\theta = 0$) and oblique ($\theta \ne 0$)
modulation cases have to be distinguished straightaway.
For small values of $k$ ($k \ll 1$), $P$ is negative and
varies as
\begin{equation}
P \bigr|_{\theta=0} \,  \approx - \frac{3}{2} \,
{\sqrt{\beta}} \,k
\end{equation}
in the parallel modulation case (i.e. $\theta = 0$), thus
tending to zero for vanishing $k$, while for $\theta \ne 0$, $P$
is positive and goes to infinity as
\begin{equation}
P \bigr|_{\theta\ne0} \,  \approx \frac{\sqrt{\beta}}{2 \, k} \,
\sin^2\theta
\end{equation}
for vanishing $k$. Therefore, the slightest deviation by $\theta$
of the amplitude variation direction with respect to the wave
propagation direction results in a change in sign of the
group-velocity dispersion coefficient $P$. On the other hand, $Q$
varies as $\sim 1/k$ for small $k \ll 1$. For $\theta \ne 0$, $Q$
is negative
\begin{equation}
Q \bigr|_{\theta\ne 0} \, \approx \, - \frac{1}{12 \,\beta^{3/2}}
\, (3 +  \beta)^2 \,
 \frac{1}{k} \, ,
 \label{lowQthetageneral}
 \end{equation}
while for vanishing $\theta$, the approximate expression for $Q$ changes
sign, i.e.
\begin{equation}
Q \bigr|_{\theta = 0} \, \approx \, + \frac{1}{12 \,\beta^{3/2}}
\, (3 + \beta)^2 \,
 \frac{1}{k} \, .
  \label{lowQthetazero}
 \end{equation}
In conclusion, both $P$ and $Q$ change sign when `switching on'
\textsl{theta}. Since the wave's (linear) stability profile,
expected to be influence by obliqueness in modulation, essentially
relies on (the sign of) the product $P Q$ (see below), we see that
long wavelengths will always be stable.

\section{Stability analysis}

The standard stability analysis [20, 21, 31] consists in
linearizing around the monochromatic (Stokes's wave) solution of
the NLSE (\ref{NLSE}): \(\psi \, = \, {\hat \psi} \, e^{i Q
|\psi|^2 \tau} \, + \, c.c. \, , \) (notice the amplitude
dependence of the frequency) by setting \({\hat \psi} \, = \,
{\hat \psi}_0 \, + \, \epsilon \, {\hat \psi}_1 \, , \) and taking
the perturbation ${\hat \psi}_1$ to be of the form: ${\hat \psi}_1
\, = \, {\hat \psi}_{1, 0} \,e^{i ({\hat k} \zeta - {\hat \omega}
\tau)} \, + \, c.c.$ (the perturbation wavenumber $\hat k$ and the
frequency $\hat \omega$ should be distinguished from their carrier
wave homologue quantities, denoted by $k$ and $\omega$).
Substituting into (\ref{NLSE}), one thus readily obtains the
nonlinear dispersion relation
\begin{equation}
\hat \omega^2 \, = \, P^2 \, \hat k^2 \, \biggl(\hat k^2 \, - \, 2
\frac{Q}{P} |\hat\psi_{0}|^2 \biggr) \, .
\end{equation}
The wave will obviously be {\em stable}
if the product $P  Q$ is negative. However, for positive $P  Q >
0$, instability sets in for wavenumbers below a critical value
$\hat k_{cr} = \sqrt{2 \frac{Q}{P}} |\hat\psi_{0}|$, i.e. for
wavelengths above a threshold: $\lambda_{cr} = 2 \pi/\hat k_{cr}$;
defining the instability growth rate \( \sigma =
|Im\hat\omega(\hat k)| \), we see that it reaches its maximum
value for $\hat k = \hat k_{cr}/\sqrt{2}$, viz.
\begin{equation} \sigma_{max} =
|Im\hat\omega|_{\hat k = \hat k_{cr}/\sqrt{2}} \,=\, | Q |\,
|\hat\psi_{0}|^2  \, . \label{growthrate}
\end{equation}
We conclude that the instability condition depends only on the
sign of the product $P Q$, which may now be studied numerically,
relying on the exact expressions derived above.

In figures \ref{figure1} to \ref{figure3}, we have depicted the $P
Q = 0$ boundary curve against the normalized wavenumber $k/k_D$
(in abscissa) and angle $\theta$ (between $0$ and $\pi$); the area
in black (white) represents the region in the $(k - \theta)$ plane
where the product is negative (positive), i.e. where the wave is
stable (unstable). For illustration purposes, we have considered w
a wide range of values of the wavenumber $k$ (normalized by the
Debye wavenumber $k_{D, h}$; nevertheless, recall that the
analysis is rigorously valid in a quite restricted region of (low)
values of $k$. Modulation angle $\theta$ is allowed to vary
between zero and ${\pi}$ (see that all plots are $\frac{\pi}{2}$-
periodic).

As analytically predicted above, the product $P Q$ is negative for
small $k$, for all values of theta; long wavelengths will always
be stable. The product possesses positive values for angle values
between zero and $\theta \approx 1$ rad $\approx 57{}^\circ$;
instability sets in above a wavenumber threshold which, even
though unrealistically high for $\theta = 0$, is clearly seen to
decrease as the modulation pitch angle $\theta$ increases from
zero to approximately 30 degrees, and then increases again up to
$\theta \approx 57{}^\circ$. Nevertheless, beyond that value (and
up to $\pi/2$) the wave remains stable; this is even true for the
wavenumber regions where the wave would be {\em unstable} to a
parallel modulation. The inverse effect is also present: even
though certain $k$ values correspond to stability for $\theta =
0$, the same modes may become unstable when subject to an oblique
modulation ($\theta \ne 0$). In all cases, the wave appears to be
globally stable to large angle $\theta$ modulation (between 1 and
$\pi/2$ radians, i.e. $57{}^\circ$ to $90{}^\circ)$.

It is interesting to trace the influence of the percentage of the
cold electron population (related to \( \beta = n_c/n_h \)) on the
qualitative remarks of the preceding paragraph. For values of
$\beta$ below unity, there seems to be only a small effect on the
wave's stability, as described above; cf. figs. \ref{figure1},
\ref{figure2}.  As a matter of fact, $\beta < 1$ appears to be
valid in most reports of satellite observations, carried out at
the edges of the Auroral Kilometric Radiation (AKR) region (where
the hot and cold electron population co-existence is observed)
\cite{Berthomier2000, obs1, obs3}; furthermore, theoretical
studies have suggested that a low $\beta$ value (in addition to a
high hot to cold electron temperature ratio) are conditions
ensuring EAW stability i.e. resistance to damping \cite{Gary,
Berthomier2000}.  Nevertheless, notice for rigor that allowing for
a high fraction of cold electrons ($\beta \gtrsim 3.5$) leads to a
strong modification of the the EA wave's stability profile, and
even produces instability in otherwise stable regions; cf. fig.
\ref{figure3} (where an unrealistic value of $\beta = 5$ was
considered). In a qualitative manner, adding cold electrons seems
to favour stability to quasi-parallel modulation (small $\theta$),
yet allows for instability to higher $\theta$ oblique modulation;
cf. figs. \ref{figure1} to \ref{figure3}. Since the black/white
regions in the figures correspond to dark/bright type solitons
(see below), we qualitatively deduce that a solitary wave of
either type may become unstable in case of an important increase
in minority electron component, i.e. well above $\beta = 1$; see
fig. \ref{figure5}. Notice that the critical value of the
cold-to-hot electron number ratio $\beta$ in order for such
phenomena to occur may be quite low if oblique modulation is
considered; see e.g. fig. \ref{figure5}b.

\section{Envelope solitary waves}

The NLSE (\ref{NLSE}) is known to possess distinct types of
localized constant profile (solitary wave) solutions, depending on
the sign of the product $P Q$. Following Ref. \cite{Hasegawa, Fedele},
we seek a solution of Eq. (\ref{NLSE}) in the form \( \psi(\zeta,
\tau) = \sqrt{\rho(\zeta, \tau)} \, e^{i\,\Theta(\zeta, \tau) }
\),  where $\rho$, $\sigma$ are real variables which are
determined by substituting into the NLSE and separating real and
imaginary parts. The different types of solution thus obtained are
summarized in the following.


For $P Q > 0$ we find the {\em (bright) envelope soliton}
\cite{commentFedele1}
\begin{equation}
\rho = \rho_0 \, sech^2\biggl(\frac{\zeta - u\, \tau}{L} \biggr)
\, , \qquad  \Theta = \frac{1}{2 P} \, \bigl[ u\,\zeta \, -
(\Omega + \frac{1}{2} u^2)\tau \bigr] \, ,
\end{equation}
representing a bell -- shaped localized pulse travelling at a speed $u$ and
oscillating at a frequency $\Omega$ (for $u = 0$). The pulse width
$L$ depends on the (constant) maximum amplitude square $\rho_0$ as
\begin{equation}
L = \sqrt{\frac{2 P}{Q \,\rho_0}} \, . \label{widthbright}
\end{equation}


For $P Q < 0$ we have the {\em dark} envelope soliton ({\em hole})
\cite{commentFedele1}
\begin{eqnarray}
\rho & = & \rho_1 \, \biggl[ 1 - \, sech^2 \biggl(\frac{\zeta -
u\, \tau}{L'} \biggr)\biggr] \, = \, \rho_1 \,
 tanh^2 \biggl(\frac{\zeta - u\, \tau}{L'}
\biggr) \, ,
\nonumber \\
\Theta & = & \frac{1}{2 P} \, \biggl[ u\,\zeta \, -
\biggl(\frac{1}{2} u^2 - 2 P Q \rho_1 \biggr) \,\tau \biggr] \, ,
\label{darksoliton}
\end{eqnarray}
representing a localized region of negative wave density (shock)
travelling at a speed $u$; this cavity traps the electron-wave envelope,
whose intensity is now rarefactive i.e.  a propagating hole in the center
and constant elsewhere.  Again, the pulse width depends on the maximum amplitude
square $\rho_1$ via
\begin{equation}
L' = \sqrt{2 \biggl|\frac{P}{Q\,\rho_1}\biggr|} \, \qquad .
\label{widthdark}
\end{equation}


Finally, looking for velocity-dependent amplitude solutions, for
$P Q < 0$, one obtains the {\em grey} envelope solitary wave
\cite{Fedele}
\begin{eqnarray}
\rho & = & \rho_2 \, \biggl[ 1 - a^2\, sech^2 \biggl(\frac{\zeta -
u\, \tau}{L''} \biggr)\biggr] \, ,
\nonumber \\
\Theta & = & \frac{1}{2 P} \, \biggl[ V_0\,\zeta \, -
\biggl(\frac{1}{2} V_0^2 - 2 P Q \rho_2 \biggr) \,\tau +
\Theta_{10} \biggr]\,
\nonumber \\ &  & \qquad \,
- S \, \sin^{-1} \frac{a\, \tanh\bigl(\frac{\zeta - u\, \tau}{L''}
\bigr)}{\biggr[  1 - a^2\, sech^2 \biggl(\frac{\zeta - u\,
\tau}{L''} \biggr) \biggr]^{1/2}} \, , \qquad \label{greysoliton}
\end{eqnarray}
which also represents a localized region of negative wave density;
$\Theta_{10}$ is a constant phase; $S$ denotes the product $S =
sign \,P \, \times sign \,(u - V_0)$. In comparison to the dark
soliton (\ref{darksoliton}), note that apart from the maximum
amplitude $\rho_2$, which is now finite (i.e. non-zero)
everywhere, the pulse width of this grey-type excitation
\begin{equation}
L'' = \sqrt{2 \biggl|\frac{P}{Q\,\rho_2}\biggr|}  \,\frac{1}{a}
\label{widthgrey}
\end{equation}
now also depends on $a$, given by
\begin{equation}
a^2 \, = \, 1 \, + \, \frac{1}{2 P Q} \frac{1}{\rho_2} (u^2 -
V_0^2) \, \le \, 1 \label{grey-depth}
\end{equation}
($P Q < 0$), an independent parameter representing the modulation
depth ($0 < a \le 1$). $V_0$ is an independent real constant which
satisfies the condition \cite{Fedele}
\[
V_0 - \sqrt{2 |P Q|\, \rho_2} \, \le \, u \, \le \,V_0 + \sqrt{2
|P Q|\, \rho_2} \quad ;
\]
for $V_0 = u$, we have $a = 1$ and thus recover the {\em dark}
soliton presented in the previous paragraph.

Summarizing, we see that the regions depicted in figs.
\ref{figure1} -- \ref{figure3} actually also distinguish the
regions where different types of localized solutions may exist:
bright (dark or grey) solitons will occur in white (black) regions
(the different types of NLS excitations are exhaustively reviewed
in \cite{Fedele}). Soliton characteristics will depend on
dispersion and nonlinearity via the $P$ and $Q$ coefficients; in
particular, the sign/absolute value of the ratio $P/Q$ provides,
as we saw, the type (bright or dark-grey)/width, respectively, of
the localized excitation.  Therefore, regions with higher values
of $|P/Q|$ will support wider (spatially more extended) localized
excitations of either bright or dark/grey type - see fig.
\ref{figure4}. Solitons of the latter type (holes) appear to be
predominant in the long wavelength region which is of interest
here, in agreement with observations, yet may become unstable and
give their place to (bright) pulses, in the presence of oblique
perturbation (figs. \ref{figure1}, \ref{figure2}) and/or local
$n_c/n_h$- value irregularities (fig. \ref{figure5}). In the short
wavelength region, these qualitative results may still be valid,
yet quantitatively appear to be rather questionable, since the wave
stability cannot be taken for granted due to electron Landau damping.
Nevertheless, even so, the EAWs are known to be less heavily
damped than Langmuir waves \cite{Gary}, and may dominate the space
plasma (high) frequency spectrum in the presence of different
temperature electron populations.

\section{Conclusions}

This work has been devoted to the study of the modulation of
EAWs propagating in an unmagnetized space plasma. Allowing for the modulation
to occur in an oblique manner, we have shown that the conditions for the
modulational instability depend on the angle between the EAW propagation
and modulation directions. In fact, the region of parameter values where
instability occurs is rather extended for angle $\theta$ values up to a certain
threshold, and, on the contrary, smeared out for higher $\theta$
values (and up to 90 degrees, then going on in a $\frac{\pi}{2}$ -
periodic fashion).

Furthermore, we have studied the possibility for the formation of
localized structures (envelope EAW solitary waves) in our two electron system.
Distinct types of localized excitations (envelope solitons) have
been shown to exist. Their type and propagation characteristics
depend on the carrier wave wavenumber $k$ and the modulation angle
$\theta$. The dominant localized mode at long wavelengths appears
to be a rarefactive region of negative wave intensity (hole),
which may however become unstable to oblique modulation or
variations of the $n_c/n_h$ ratio. It should be mentioned that
both bright and dark/grey envelope excitations are possible within
this model; thus, even though the latter appear to be rather
privileged within the parameter range where waves are expected not
to be heavily damped, the former may exist due to oblique
amplitude perturbations. In conclusion, we stress that the
qualitative aspects of the observed envelope solitary structures
are recovered from our simple fluid model. The present investigation
can be readily extended to include the effects of the geomagnetic
field, a tenuous electron beam, and on dynamics on the amplitude
modulation of the EAWs. The magnetic field effects would
produce three-dimensional NLSE in which the longitudinal and
transverse (to the external magnetic field direction) group dispersions
would be different due to the cold electron polarization effect.
The harmonic generation nonlinearities would also be modified by
the presence of the external magnetic field.

\bigskip

\begin{acknowledgments}
This work was supported by the European Commission (Brussels)
through the Human Potential Research and Training Network for
carrying out the task of the project entitled: ``Complex Plasmas:
The Science of Laboratory Colloidal Plasmas and Mesospheric
Charged Aerosols'' through the Contract No. HPRN-CT-2000-00140.
\end{acknowledgments}


\newpage


\centerline{\textbf{Figure Captions}}

Figure 1.

The product $P Q = 0$ contour is depicted against the normalized
wavenumber $k/k_D$ (in abscissa) and angle $\theta$ (between $0$
and $\pi$); black (white) colour represents the region where the
product is negative (positive), i.e. the region of linear
stability (instability). Furthermore, black (white) regions may
support dark (bright)-type solitary excitations. This plot refers
to a realistic cold to hot electron ratio equal to
$\beta = 0.5$ (i.e. one third of the electrons are cold).

\bigskip

Figure 2.

\begin{figure}[htb]
 \centering
 \resizebox{3in}{!}{
 \includegraphics[]{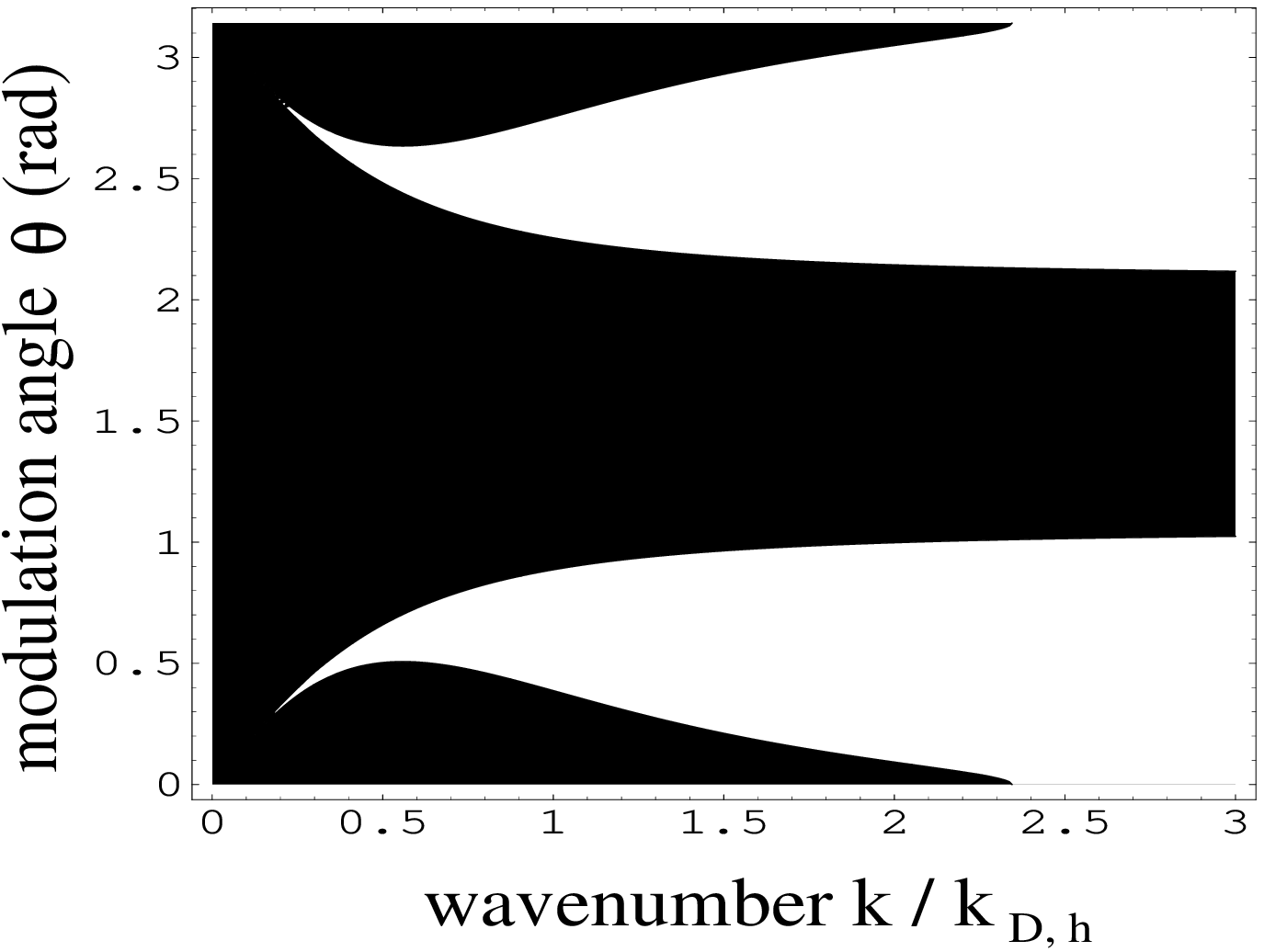}}
 \caption{} \label{figure1}
\end{figure}

Similar to fig. \ref{figure1}, for $\beta = 1$.

\bigskip

Figure 3.

\begin{figure}[htb]
 \centering
 \resizebox{3in}{!}{
 \includegraphics[]{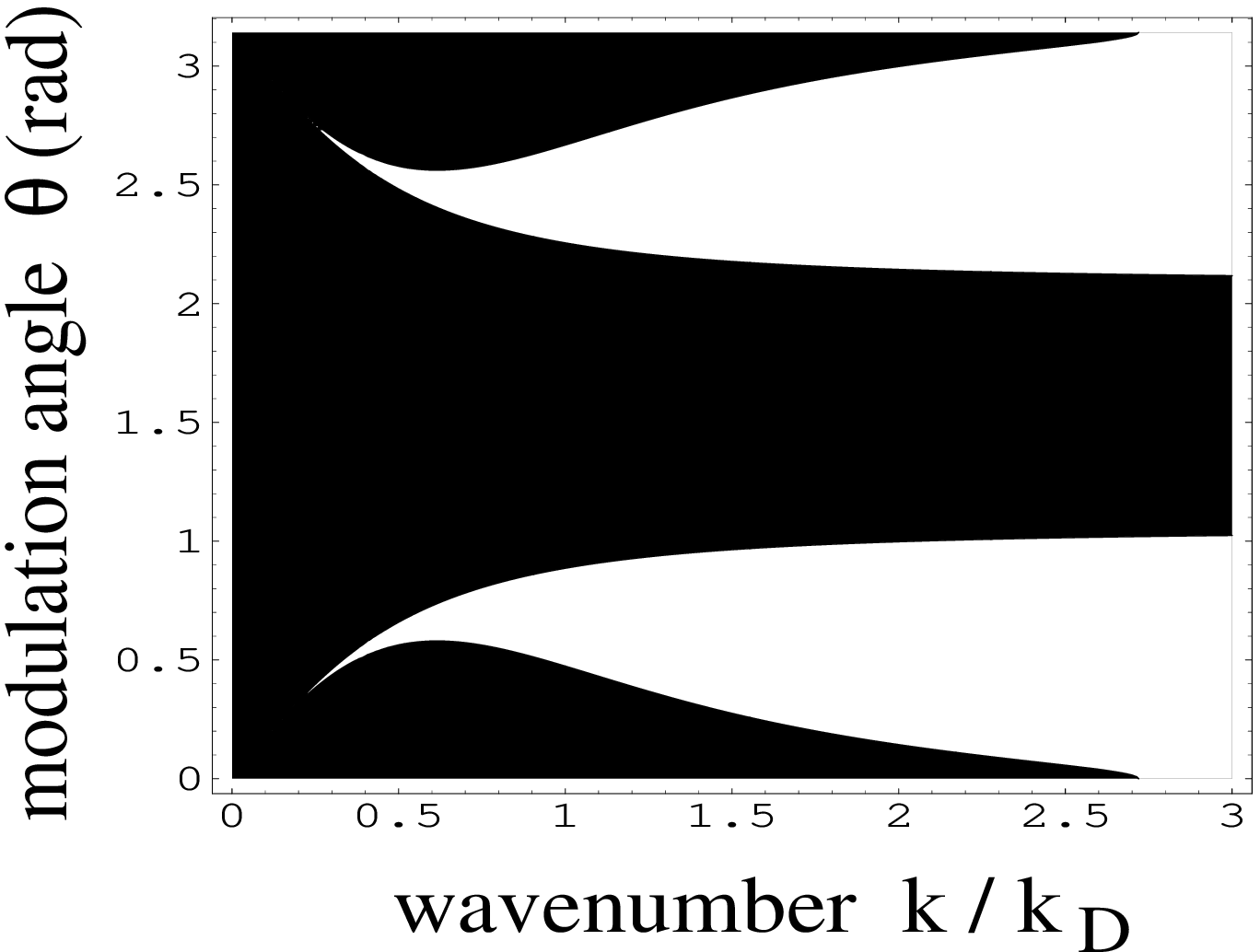}}
 \caption{} \label{figure2}
\end{figure}

\begin{figure}[htb]
 \centering
 \resizebox{3in}{!}{
 \includegraphics[]{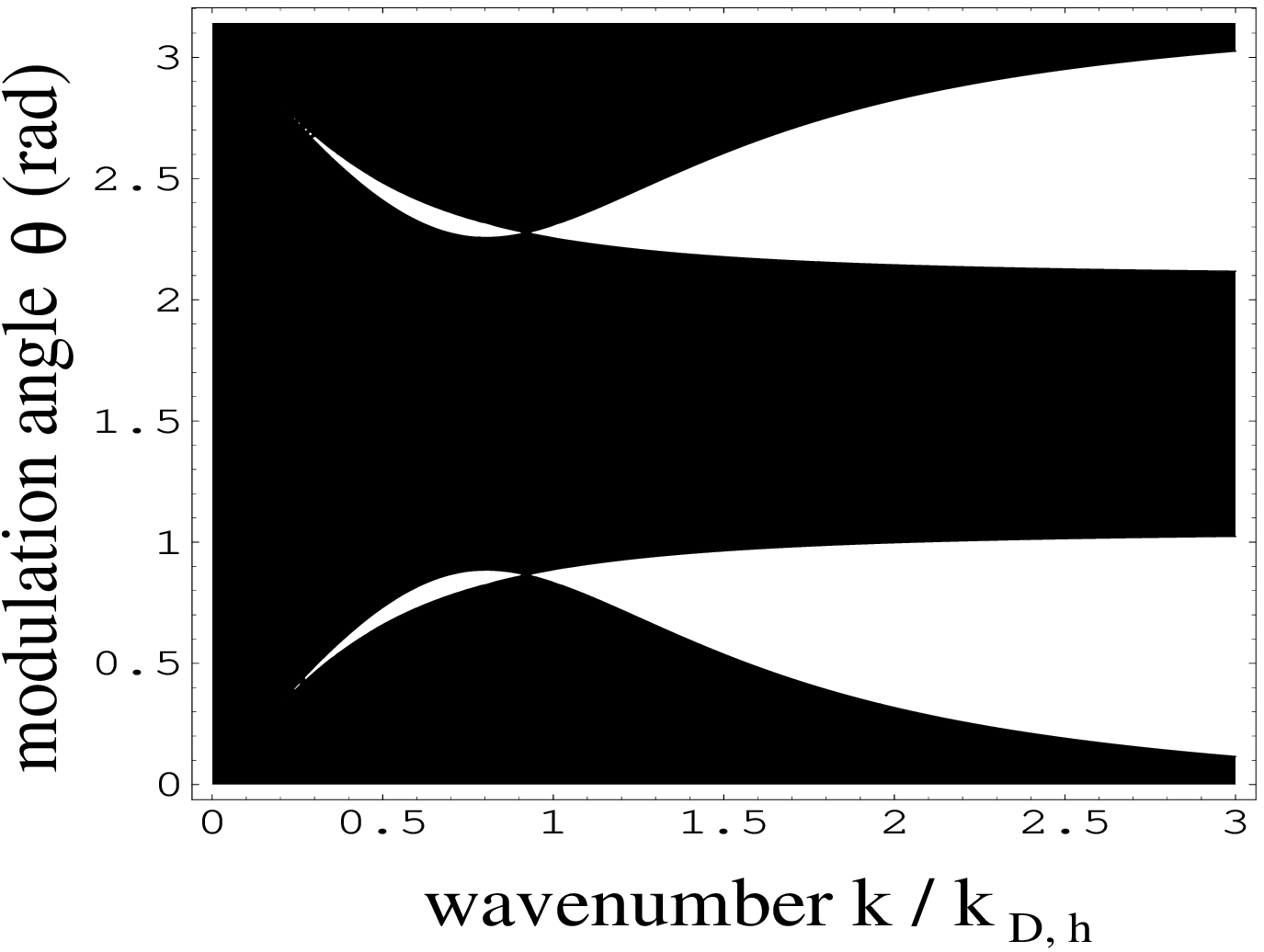}}
 \caption{} \label{figure3}
\end{figure}

\begin{figure}[htb]
 \centering
 \resizebox{3in}{!}{
 \includegraphics[]{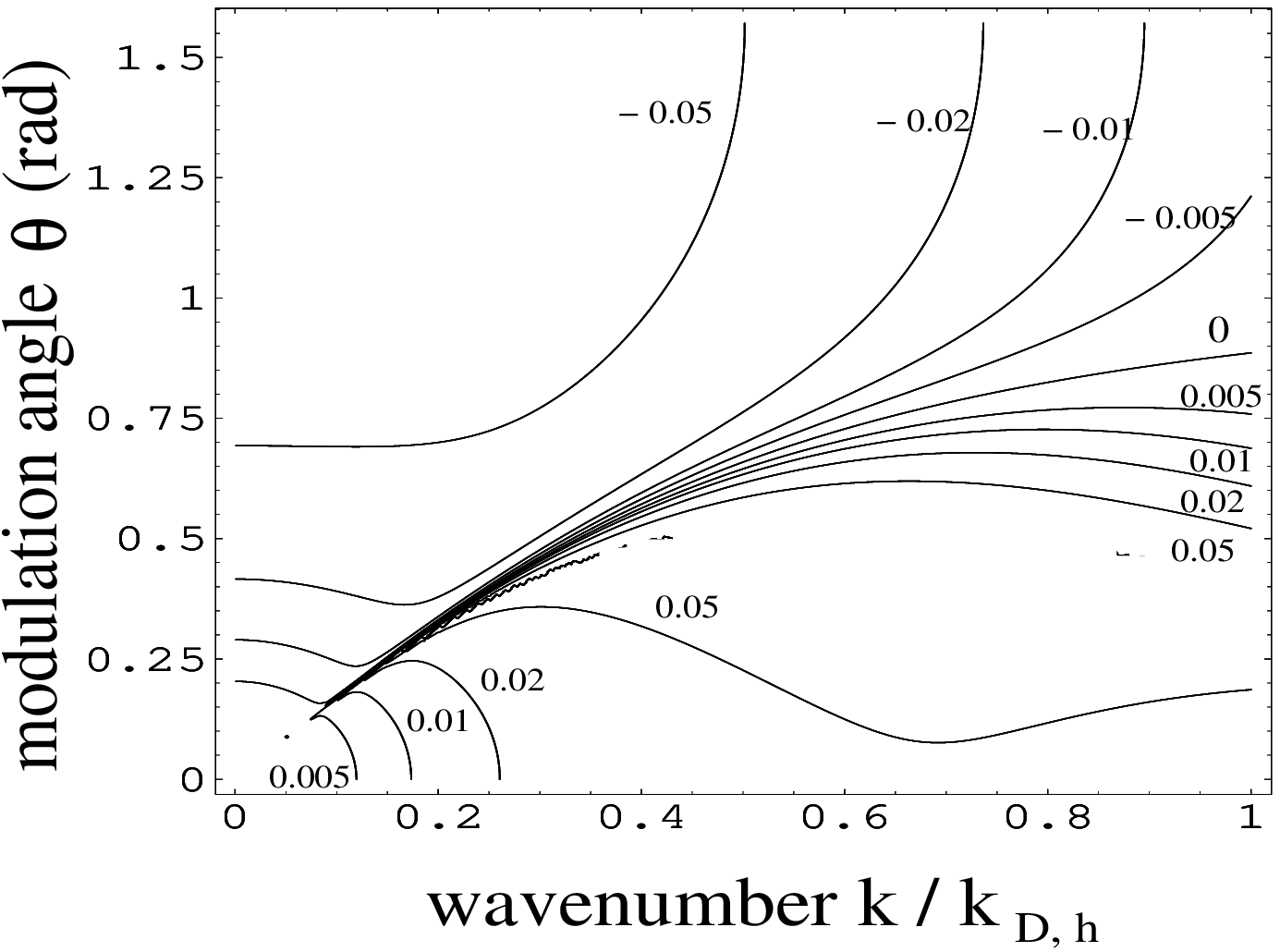}}
 \caption{} \label{figure4}
\end{figure}

\begin{figure}[htb]
 \centering
 \resizebox{3in}{!}{
 \includegraphics[]{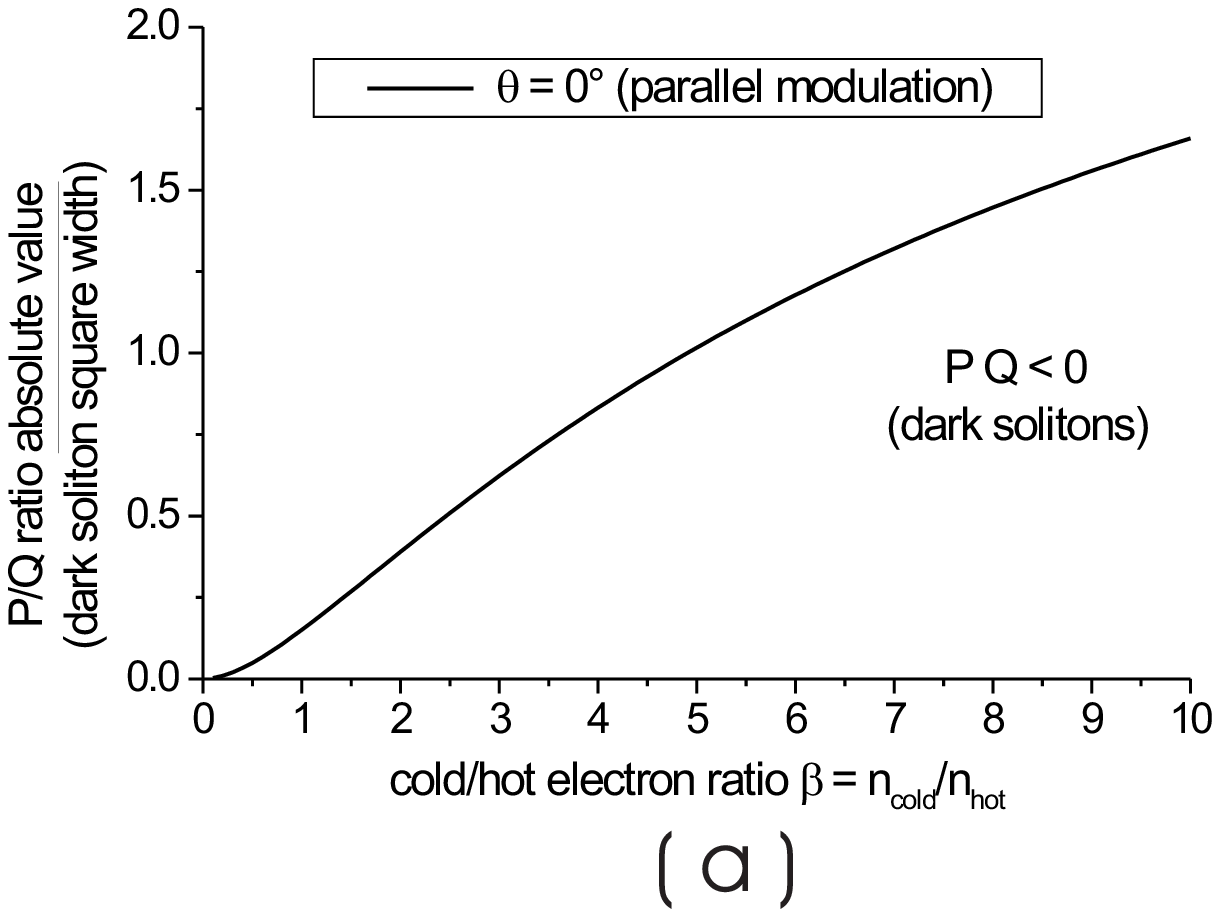}}
 \\
 \vskip 2 cm
 \centering
 \resizebox{3in}{!}{
 \includegraphics[]{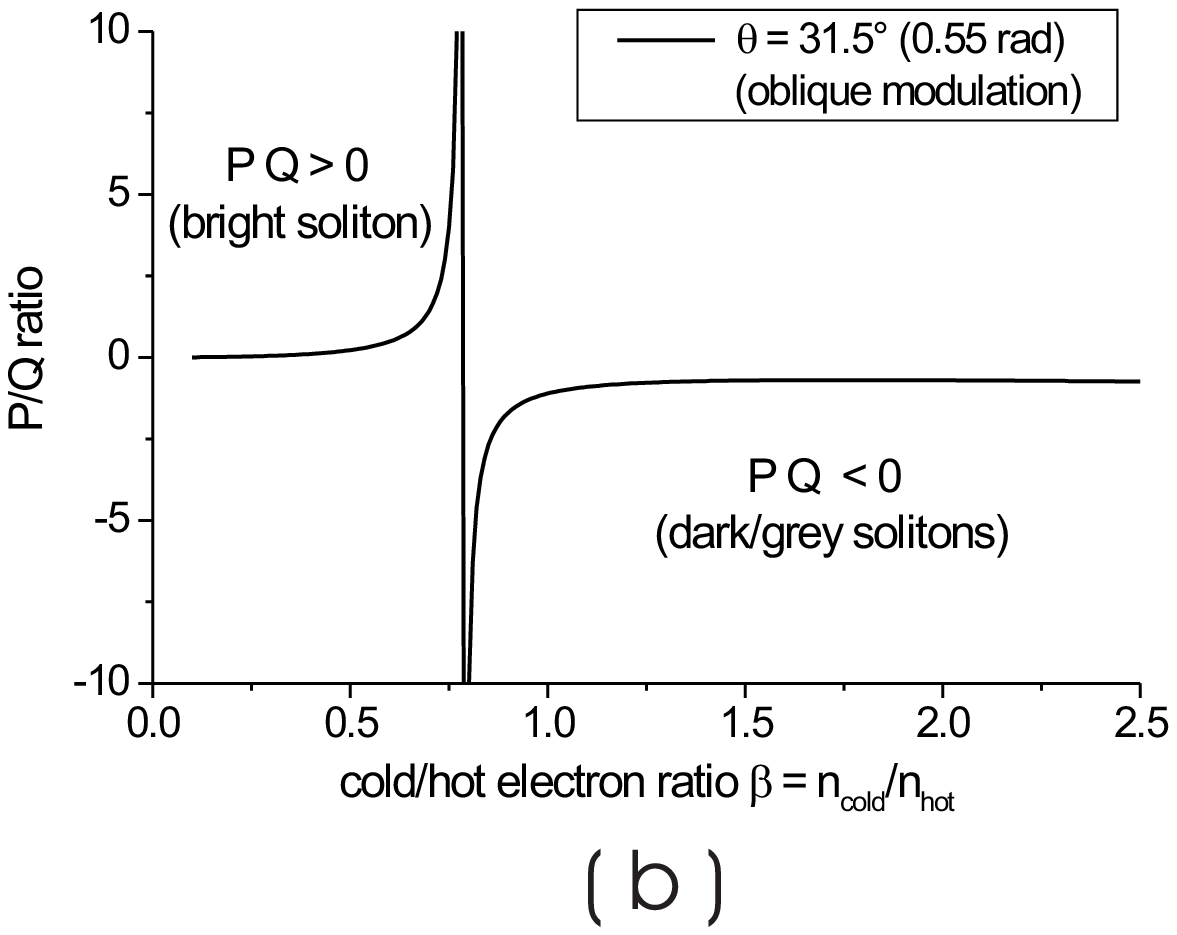}}
  \\
   \vskip 2 cm
 \centering
 \resizebox{3in}{!}{
 \includegraphics[]{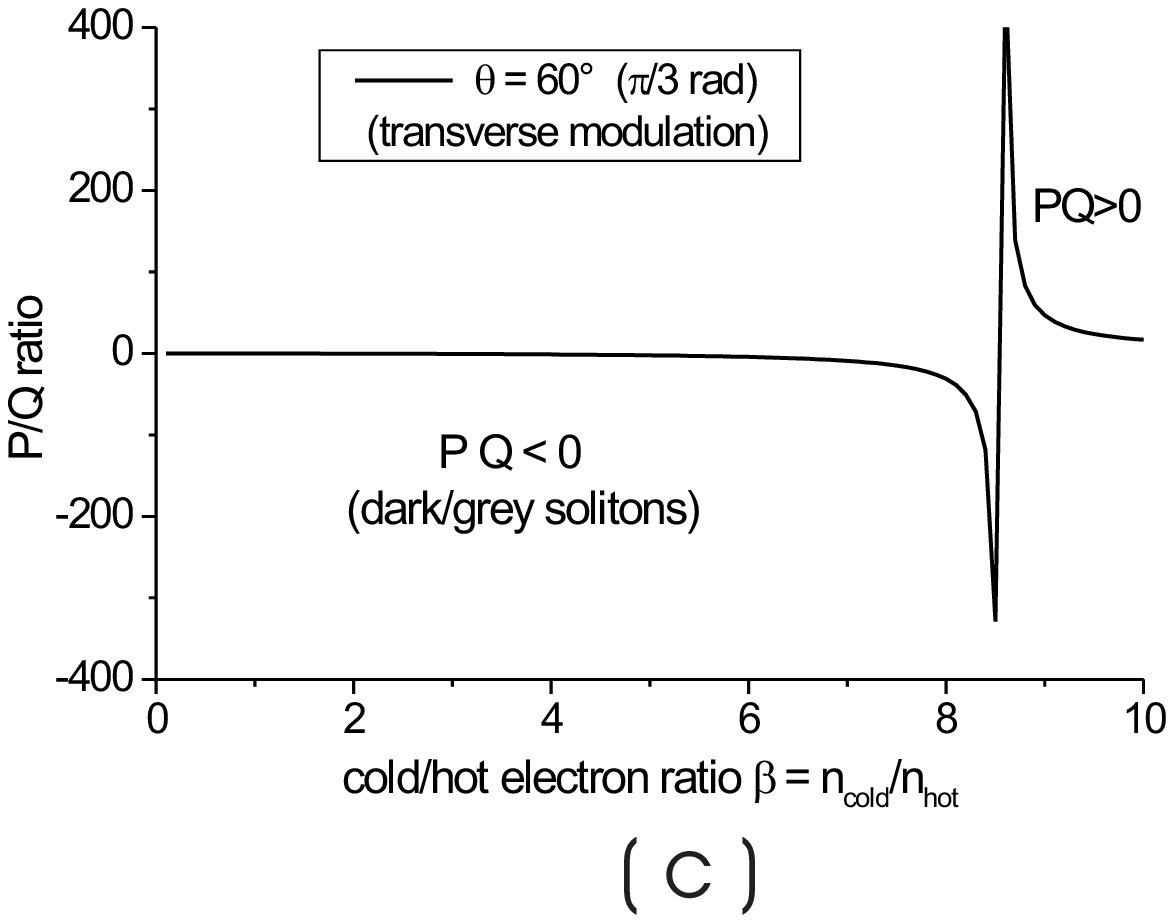}}
 \caption{} \label{figure5}
\end{figure}

Similar to figures \ref{figure1}, \ref{figure2} considering a very
strong presence of cold electrons ($\beta = 5$). Notice the
appearance of instability (bright) regions for large angle values
and long wavelengths.

\bigskip

Figure 4.

Contours of the ratio $P/Q$ -- whose absolute value is related to
the square of the soliton width, see (\ref{widthbright}),
(\ref{widthdark}) -- are represented against the normalized
wavenumber $k/k_{D, h}$ and angle $\theta$. See that the negative
values correspond to two branches (lower half), so that the
variation of $P/Q$, for a given wavenumber $k$, does not depend
monotonically on $\theta$. $\beta = 0.5$ in this plot.

\bigskip

Figure 5.

The $P/Q$ coefficient ratio, whose sign/absolute value is related
to the type/width of solitary excitations, is depicted against the
cold-to-hot electron density ratio $\beta$. The wavenumber is chosen
as $k/k_{D, h} = 0.7$. (a) $\theta = 0{}^\circ$
(parallel modulation): only dark-type excitations exist ($P Q <
0$); their width increases with $\beta$. (b) $\theta = 60{}^\circ$
(oblique modulation): bright/dark excitations exist for $\beta$
below/above $\beta_{cr} \approx 0.8$. The bright/dark soliton
width increases/decreases with $\beta$. (c) $\theta = 90{}^\circ$
(transverse modulation). A rather (unacceptably) high value of
$\beta$ was taken, to stress the omnipresence of dark--type
excitations.

\newpage

\vskip 2 cm

\newpage

\vskip 2 cm

\newpage

\vskip 2 cm

\newpage

\vskip 2 cm

\newpage

\vskip 2 cm

\begin{thebibliography}{10}
\bibitem{commentFedele1} This result is immediately obtained from Ref.
\cite{Fedele}, by transforming the variables therein into our
notation as follows: $x \rightarrow \zeta$, \, $s \rightarrow
\tau$, \, $\rho_m \rightarrow \rho_0$, \ $\alpha \rightarrow 2 P$,
\ $q_0 \rightarrow - 2 P Q$, \ $\Delta \rightarrow L$, \ $E
\rightarrow \Omega$, \ $V_0 \rightarrow u$.
\end{thebibliography}
\end{document}